\documentclass[twocolumn,showpacs,preprintnumbers,amsmath,amssymb]{revtex4}
\usepackage{dcolumn}% Align table columns on decimal point
\usepackage{bm}% bold math
\usepackage{ae} % or {zefonts}
\usepackage[T1]{fontenc}
\usepackage[ansinew]{inputenc}
\usepackage{amsmath}
\usepackage{graphicx}
\usepackage{color}
\usepackage[colorlinks]{hyperref}

\begin{document}

\preprint{APS/123-QED}

\title{Asymmetric phase diagram of mixed CuInP$_2$(S$_x$Se$_{1-x}$)$_6$  crystals }% Force line breaks with \\

\author{J.Macutkevic$^1$}
\author{J.Banys$^2$}
\email{juras.banys@ff.vu.lt}
%\altaffiliation
\author{R. Grigalaitis$^2$}
\author{Yu. Vysochanskii$^3$}
\affiliation{$^1$Semiconductor Physics Institute, A. Gostauto 11,
2600 Vilnius, Lithuania}
\affiliation{$^2$Faculty of Physics, Vilnius University,
Sauletekio 9, Vilnius LT-10222, Lithuania}
\affiliation{$^3$Institute of Solid State
Physics and Chemistry of Uzhgorod University, Ukraine}

\date{\today}% It is always \today, today,

\begin{abstract}
In this article mixed CuInP$_2$(S$_x$Se$_{1-x}$)$_6$ crystals were
investigated by broadband dielectric spectroscopy (20 Hz - 3 GHz).
The complete phase diagram has been obtained from these results. The phase diagram of investigated crystals is
strongly asymmetric - the decreasing of ferroelectric phase
transition temperatures in CuInP$_2$(S$_x$Se$_{1-x}$)$_6$ is much
more flat with small admixture of sulphur then with small
admixture of selenium. In the middle part of the phase diagram (x=0.4-0.9) the dipolar glass phase
has been observed. In boundary region between ferroelectric
order and dipolar glass disorder with small amount of sulphur (x=0.2-0.25) at low temperatures the
nonergodic relaxor phase appears. The phase diagram was discussed in terms of random bonds
and random fields.

\end{abstract}

\maketitle

\section{\label{sec:level1}Introduction }
Solid systems present many interesting types of phase transitions, with ferro, antiferro, or modulated
long range order at lower temperatures. Disordered cooperative systems have also
attracted a lot of attention. Nonergodic relaxor, dipolar glass phases or coexistence of
ferroelectric and dipolar glass phases can appear in disordered systems at low temperatures. The nature of these phases continues to generate considerably
experimental and theoretical interest.

CuInP$_2$S$_6$ crystals represent an unusual example of a
anticollinear two-sublattice ferrielectric system \cite{maisonneuve1,
bourdon1, cajipe1, vysochanskii1}.  Here a first-order phase
transition of the order-disorder type from the paraelectric to the
ferrielectric phase is realized (\textit{T$_c$} = 315 K). The symmetry
reduction at the phase transition (C2/c $\rightarrow$ Cc) occurs
due to ordering in the copper sublattice and displacement of
cations from the centrosymmetric positions in the indium
sublattice.  The spontaneous polarization arising at the phase
transition to the ferrielectric phase is perpendicular to the
layer planes. These thiophosphates consist of lamellae defined by
a sulphur framework in which the metal cations and P - P pairs
fill the octahedral voids; within a layer, the Cu, In, and P-P
form triangular patterns \cite{maisonneuve1, bourdon1, cajipe1}.
The cation off-centering, 1.6 Å for Cu$^I$ and 0.2 Å for
In$^{III}$, may be attributed to a second-order Jahn-Teller
instability associated with the \textit{d$^{10}$} electronic configuration.
The lamellar matrix absorbs the structural deformations via the
flexible P$_2$S$_6$ groups while restricting the cations to
antiparallel displacements that minimize the energy costs of
dipole ordering. Each Cu ion can occupy two different positions.
The Cu, In and P - P form triangular patterns within the layer.
Relaxational rather than resonant behaviour is indicated by the
temperature dependence of the spectral characteristics, is in
agreement with X$-$ray investigations. It was suggested that a
coupling between P$_2$S$_6$ deformation modes and Cu$^I$
vibrations enables the copper ion hopping motions that lead to the
loss of polarity and the onset of ionic conductivity in this
material at higher temperatures \cite{vysochanskii1}. The investigation of ionic
conductivity in CuInP$_2$S$_6$ \cite{maisonneuve2, banys1} have showed that $\sigma_{DC}$
follows the Arrhenius law with the activation energy \textit{E$_A$} = 0.73
eV \cite{maisonneuve2} and more detailed investigations showed
\textit{E$_A$}=0.635 eV \cite{banys1}.

The results of dielectric investigations  of CuInP$_2$Se$_6$
showed two phase transition:  a second-order one at \textit{T$_i$}= 248 K
and a first-order transition at \textit{T$_c$} =236 K \cite{vysochanskii2}.
The results followed to the conclusion that an incommensurate phase occurs
between \textit{T$_i$} and \textit{T$_c$}. However, the calorimetric investigations
showed only a broad phase transition between 220 and 240 K in this
compound \cite{bourdon2}. More accurate broadband dielectric
investigations showed only nearly second order phase transition at
\textit{T$_c$}=226 K \cite{samulionis}. From a single-crystal X-ray
diffraction study follows that the high- and low-temperature
structures of CuInP$_2$Se$_6$ (trigonal space group P-31c and
P31c, respectively) are very similar to those of CuInP$_2$S$_6$ in
the paraelectric and ferrielectric phases, with the Cu$^I$
off-centering shift being smaller in the former than in the latter
\cite{bourdon1,bourdon2}. There the thermal evolution of the cell
parameters of CuInP$_2$Se$_6$ was obtained by full profile fits to
the X-ray diffractograms. Both cell parameters \textit{a} and
\textit{c} slightly decrease on cooling, and \textit{a} parameter shows a local minimum at \textit{T}=226 K. This behaviour is
quite different from the anomalous increases found in the cell
parameters of CuInP$_2$S$_6$ when heating through the transition
\cite{maisonneuve1, bourdon1}.

The important feature of selenides is the higher covalence degree
of their bonds. Evidently, for this reason the copper ion sites in
the low-temperature phase of CuInP$_2$Se$_6$ are displaced only by
1.17 Å \cite{bourdon2} from the middle of the structure layers  in
comparison with the corresponding displacement 1.6 Å for
CuInP$_2$S$_6$ \cite{maisonneuve1}. These facts enable to assume
that the potential relief for copper ions in CuInP$_2$Se$_6$ is
shallower than for its sulphide analog. Presumably, for this
reason the structural phase transition in the selenide compound is
observed at lower temperature than for the sulphide compound.
Preliminary dielectric investigations of
CuInP$_2$(S$_x$Se$_{1-x}$)$_6$ crystals are presented in
\cite{banys2,vysochanskii4}. The data in \cite{vysochanskii4} is
measured only at frequency 10 kHz and the paper \cite{banys2}
contain data only on one compound -
CuInP$_2$(S$_{0.7}$Se$_{0.3}$)$_6$.

The aim of this paper is to investigate phase diagram of mixed
CuInP$_2$(S$_x$Se$_{1-x}$)$_6$ crystals via broadband dielectric
spectroscopy. We showed that in mixed crystals
with the increasing amount of impurities two smearing of
ferroelectric phase transition scenarios are possible:
ferroelectric - inhomogeneous ferroelectric - dipolar glass or
ferroelectric - relaxor - dipolar glass.

\section{\label{sec:level1}Experimental }
Crystals of CuInP$_2$(S$_x$Se$_{1-x}$)$_6$ were grown by Bridgman
method. For the dielectric spectroscopy the plate like crystals
were used. All measurements were performed in direction
perpendicular to the layers. The complex dielectric permittivity
$\varepsilon$$^*$ was measured using the HP4284A capacitance
bridge in the frequency range 20 Hz to 1 MHz. In the frequency
region from 1 MHz to 3 GHz measurements were performed by a
coaxial dielectric spectrometer with vector network analyzer
Agilent 8714ET. All measurements have been performed on cooling
with controlled temperature rate 0.25 K/min. Silver paste has been used for
contacting.
\section{\label{sec:level2} Results and discussion}
\subsection{\label{sec:levelA} Influence of small amount
of sulphur to phase transition dynamics in CuInP$_2$Se$_6$
crystals} A small amount of admixture can significant changes
properties of ferroelectrics. In mixed
CuInP$_2$(S$_x$Se$_{1-x}$)$_6$ crystals with \textit{x}$\leq$0.1 the
ferroelectric phase transition is observed (Fig. 1). Here the dielectric permittivity maximum
temperature (\textit{T$_m$}) is frequency-dependent only at higher frequencies (above 1 MHz). The phase transition temperature can be defined by \textit{T$_m$} at low frequencies (below 1 MHz). The temperature behaviour of the dielectric dispersion of CuInP$_2$Se$_6$ crystals with a small admixture of sulphur (Fig. 2) is very similar to the dielectric dispersion of pure CuInP$_2$Se$_6$  crystals \cite{samulionis}. At higher temperatures (T ${>>}$ T$_c$) the dielectric dispersion reveals in 10$^8$ - 10$^{10}$ Hz frequency range. With decreasing temperature the dielectric dispersion become broader and appears at lower frequencies. At lower temperatures (T  ${<<}$  T$_c$) the dielectric dispersion remains in the 10$^6$ - 10$^{10}$ Hz frequency range and only its strength decreases on cooling.  
\begin{figure} [b]
  \begin{center}
    \includegraphics[width=83mm]{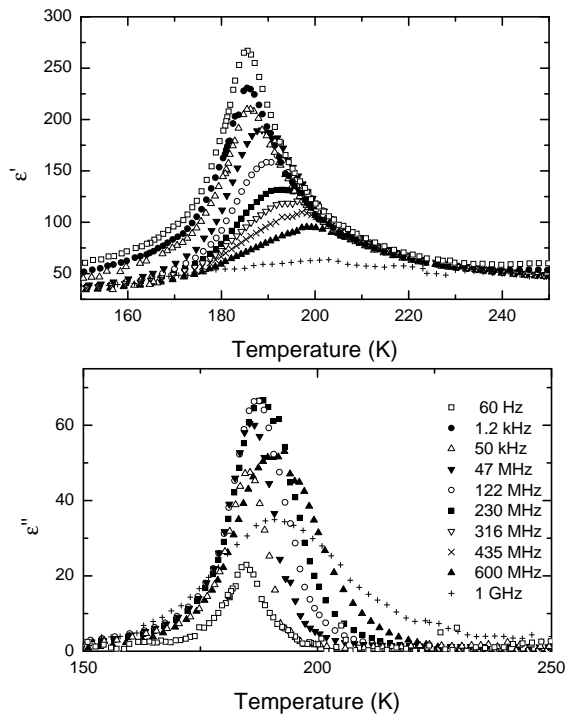}
  \end{center}
    \caption{
    Temperature dependence of the complex dielectric permittivity of CuInP$_2$(S$_{0.1}$Se$_{0.9}$)$_6$
    crystals measured at several
frequencies. }
    \label{Fig_1}
\end{figure}
 \begin{figure} [b]
  \begin{center}
    \includegraphics[width=83mm]{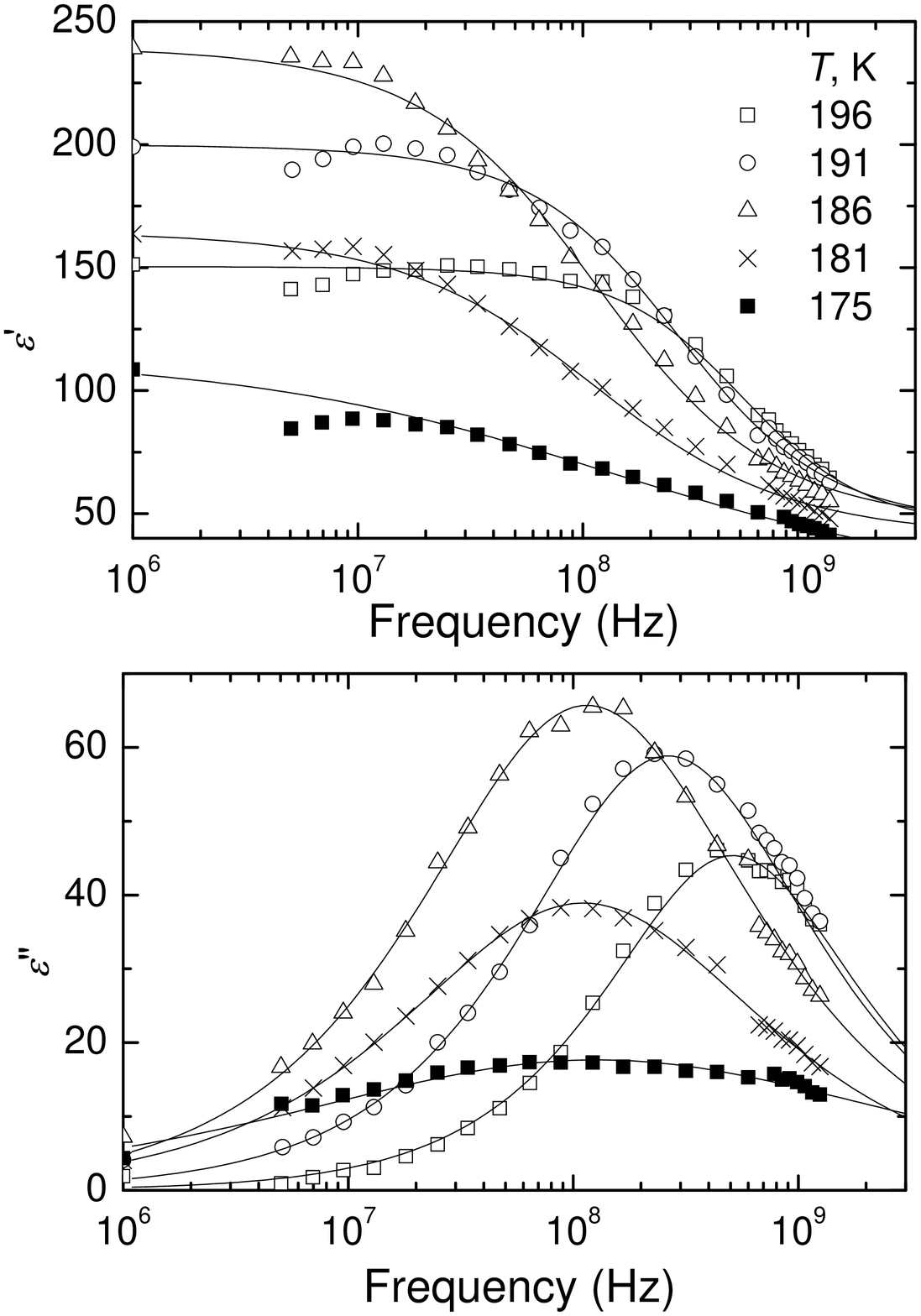}
  \end{center}
    \caption{
    Frequency dependence of the complex dielectric permittivity of CuInP$_2$(S$_{0.1}$Se$_{0.9}$)$_6$
    crystals measured at several
temperatures. Lines are results of Cole-Cole fits.}
    \label{Fig_2}
\end{figure}
More information about the phase transition dynamics can be obtained by
analysis the dielectric dispersion with the Cole-Cole formula
\begin{equation}\label{Cole}
\varepsilon^{*}(\nu)=\varepsilon_{\infty}+\frac{\Delta\varepsilon}{1+(i\omega\tau_{CC})^{1-\alpha_{CC}}},
\end{equation}
where $\Delta$$\varepsilon$ represents dielectric strength of the
relaxation, $\tau_{CC}$ is the mean Cole-Cole relaxation time,
$\varepsilon$$_{\infty}$ represents the contribution of all polar
phonons and electronic polarization to the dielectric permittivity
and $\alpha$$_{CC}$ is the Cole-Cole relaxation time distribution
parameter; when $\alpha$$_{CC}$=0, Eq.~\ref{Cole} reduces to the
Debye formula. Obtained parameters are presented in Fig. 3.
The Cole-Cole parameters of all presented compounds show the
similar behaviour: the Cole-Cole distribution parameter
$\alpha$$_{CC}$ strongly increases on cooling, reciprocal
dielectric strength 1/$\Delta$$\varepsilon$ exhibits a minimum at
ferroelectric phase transition temperature, the soft mode
frequency $\nu$$_{r}$= 1/(2$\pi$$\tau$$_{CC}$) slows down on
cooling in the paraelectric phase. The temperature dependence of
the dielectric strength $\Delta$$\varepsilon$  was fitted with the
Curie-Weiss law (Fig. 3)
\begin{equation}\label{Ciuri}
\Delta\varepsilon=C_{p,f}/(|T-T_C|),
\end{equation}
where \textit{C$_{p,f}$} is the Curie-Weiss constant and \textit{T$_C$} is the
Curie-Weiss temperature. The temperature dependence of soft mode
frequency $\nu$$_r$ in paraelectric phase was fitted with the equation
\begin{equation}\label{clasic}
\nu_r=A(T-T_C),
\end{equation}
where \textit{A} is a constant. Obtained parameters are presented in Table 1.
The phase transition temperature \textit{T$_C$}  in mixed crystals strongly
decreases from 225 K to 185 K. For all the compounds the
\textit{C$_p$}/\textit{C$_f$} ratio is about 1.5, for the second order phase
transitions this ratio must be 2, for the first order one - higher
than 2. The assumption was made that in these crystals between
paraelectric and ferroelectric phase an additional incommensurate
phase exists \cite{vysochanskii4}. However, in all mixed
CuInP$_2$(S$_x$Se$_{1-x}$)$_6$ crystals with \textit{x}$\leq$0.1 no
anomaly above the main (ferroelectric) phase transition was
observed (Fig. 1).

\begin{table}
\caption{\label{table1} Parameters of phase transition dynamic of
CuInP$_2$Se$_6$ crystals with small admixture of sulphur (\textit{x}$\leq$0.1).}
\begin{ruledtabular}
\begin{tabular}{ccccc}
  compound & \textit{C$_p$}, K & \textit{C$_p$}/\textit{C$_f$} & \textit{A}, MHz/K & \textit{T$_C$}, K \\
\hline

\hline\\

CuInP$_2$Se$_6$ from \cite{samulionis}  & 591.7 & 1.33 & 271.9 & 225 \\
CuInP$_2$(Se$_{0.98}$S$_{0.02}$)$_6$  & 309.6 & 1.43 & 193.4 & 215.7 \\
CuInP$_2$(Se$_{0.95}$S$_{0.05}$)$_6$  & 980.3 & 1.66 & 79.3 & 208.2 \\
CuInP$_2$(Se$_{0.9}$S$_{0.1}$)$_6$  & 2380.9 & 1.52 & 44.4 & 185 \\
           \end{tabular}
 \end{ruledtabular}
\end{table}
\begin{figure} [t]
\begin{center}
    \includegraphics[width=80mm]{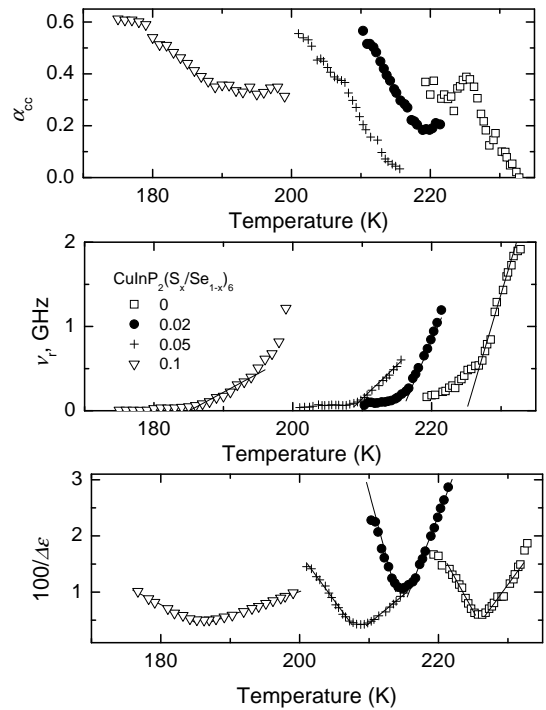}
  \end{center}
\caption{ Temperature dependence of the Cole-Cole parameters of
complex dielectric permittivity for the
CuInP$_2$(S$_{x}$Se$_{1-x}$)$_6$
    crystals with \textit{x}$\leq$0.1. The
 $\nu$$_{r}$ lines were obtained from fit with Eq.~\ref{clasic} and the 1/$\Delta$$\varepsilon$ lines were obtained from Curie-Weiss
fit. The data for CuInP$_2$Se$_6$ is from \cite{samulionis}.}
\label{Fig_3}
\end{figure}
Below the ferroelectric phase transition temperature the  dielectric dispersion is broad and part of it
appears in the low frequency region (Fig. 1). This part is caused by
ferroelectric domain dynamics. Therefore, the contribution of
ferroelectric domain dynamics effectively raises the dielectric
strength $\Delta$$\varepsilon$ in the ferroelectric phase and
\textit{C$_f$} constant.

\subsection{\label{sec:levelB} Nonergodic relaxor phase in mixed CuInP$_2$(S$_x$Se$_{1-x}$)$_6$ crystals}
Recently, the relaxor-like behaviour as an embryo of the glass
state is proposed in the antiferroelectric-glass phase boundary
region of DRADP crystals family \cite{matsuhita1}. Here it is showed that the growth of glass ordering is in quite a different pattern from that of the
ferroelectric-glass phase boundary region. In
this section we shall presented two very similar
CuInP$_2$(S$_x$Se$_{1-x}$)$_6$ compounds (x=0.2 and x=0.25), which
exhibit peculiar dielectric behaviour. Each composition shows just
one maximum in $\varepsilon'$(T) and $\varepsilon''$(T) in the
range of 110 and 145\,K at frequency 10 kHz \cite{vysochanskii4}.
\begin{figure}[h]
 \begin{center}
    \includegraphics[width=83mm]{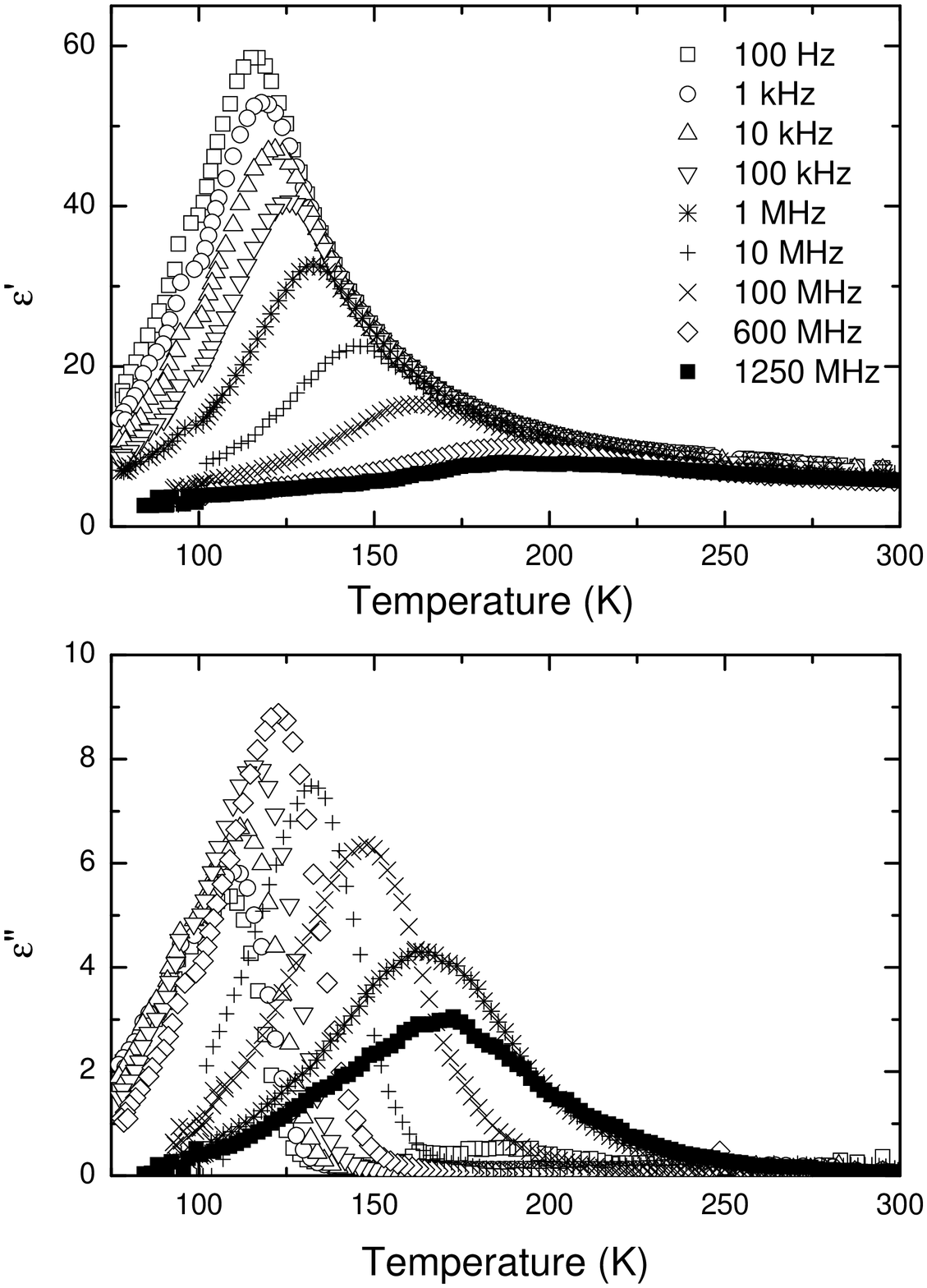}
  \end{center}
   \caption{
    Temperature dependence of the complex dielectric permittivity of CuInP$_2$(S$_{0.25}$Se$_{0.75}$)$_6$
    crystals measured at several
frequencies. }
    \label{Fig_4}
\end{figure}
The temperature dependences of the complex dielectric permittivity
${\varepsilon^{*}}$ at various frequencies of these crystals show
typical relaxor behaviour. As an example, dielectric permittivity
of CuInP$_2$(Se$_{0.75}$S$_{0.25}$)$_6$ crystal is shown in Fig. 4.
There is a broad peak in the real part of dielectric permittivity is observed.
With frequency \textit{T$_m$} and the magnitude of the peak increases in the whole frequency range.
\begin{figure}
\begin{center}
    \includegraphics[width=83mm]{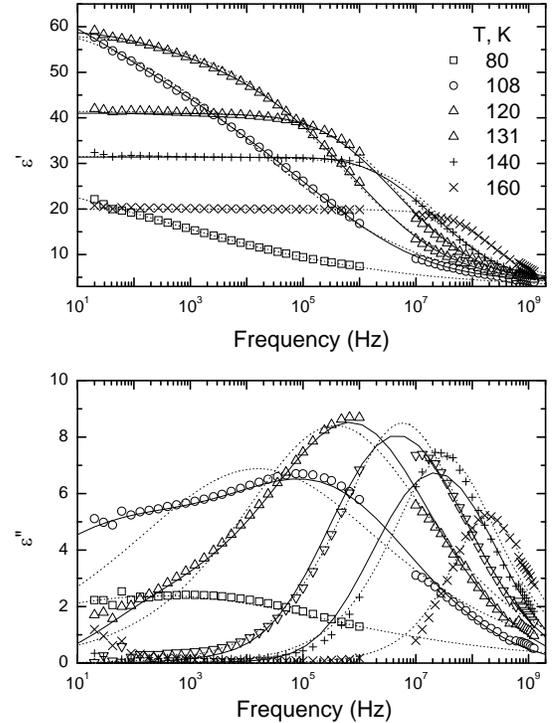}
  \end{center}
\caption{ Frequency dependence of the complex dielectric
permittivity of CuInP$_2$(S$_{0.25}$Se$_{0.75}$)$_6$
    crystals at several
temperatures. Lines are results of fits with distributions of
relaxation times (solid) and of Cole-Cole fit (dot). }
\label{Fig_5}
\end{figure}
There is a strong dielectric
dispersion in a radio frequency region around and below \textit{T$_m$} at 1
kHz. The value of \textit{T$_{mm}$} (the temperature of the maximum of
losses) is much lower than that of \textit{T$_m$} at the same frequency.
The position of the maximum of dielectric permittivity is strongly
frequency-dependent; no certain static dielectric permittivity can
be obtained below and around dielectric permittivity maximum
temperature \textit{T$_m$} at 1 kHz. Such behaviour can be described by the
Vogel-Fulcher relationship
\begin{equation}\label{Vogel-Fulcher}
\nu=\nu_{0}\exp{\frac{E_{f}}{k(T_{m}-T_{0t})}},
\end{equation}
where \textit{k} is the Boltzman constant, \textit{E$_f$}, ${\nu}$$_{0}$, \textit{T$_{0t}$} are parameters of this equations. Obtained parameters are presented in Table~\ref{table2}.

\begin{table}
\caption{\label{table2} Parameters of the Vogel-Fulcher fit of the
 \textit{T$_m$} dependence of frequency for
CuInP$_2$(S$_x$Se$_{1-x}$)$_6$ crystals with 0.2$\leq$\textit{x}$\leq$0.25. }
\begin{ruledtabular}
\begin{tabular}{ccccc}
  compound & ${\nu}$$_0$, GHz & \textit{T$_{0t}$}, K & \textit{E$_f$}/\textit{k}, K \\
\hline

\hline\\

  CuInP$_2$(Se$_{0.75}$S$_{0.25}$)$_6$  & 38.34 & 96.8  & 370 \\
  CuInP$_2$(Se$_{0.8}$S$_{0.2}$)$_6$  & 10.96 & 134.5 & 150 \\
           \end{tabular}
 \end{ruledtabular}
\end{table}
The dielectric dispersion of CuInP$_2$(Se$_{0.75}$S$_{0.25}$)$_6$
crystals show strong temperature dependence (Fig. 5): at higher
temperatures the dielectric dispersion is only in 10$^7$ -
10$^{10}$ Hz region, on cooling the dielectric dispersion becomes
broader and more asymmetric. Strongly asymmetric and very broad
dielectric dispersion is observed below dielectric permittivity
maximum temperature \textit{T$_m$} at 1 kHz. The Cole-Cole formula (Eq.~\ref{Cole}) can describe such dielectric dispersion only at
higher temperatures, due to predefined symmetric shape of the
distribution of the relaxations times.
\begin{figure}
\begin{center}
    \includegraphics[width=83mm,height=83mm]{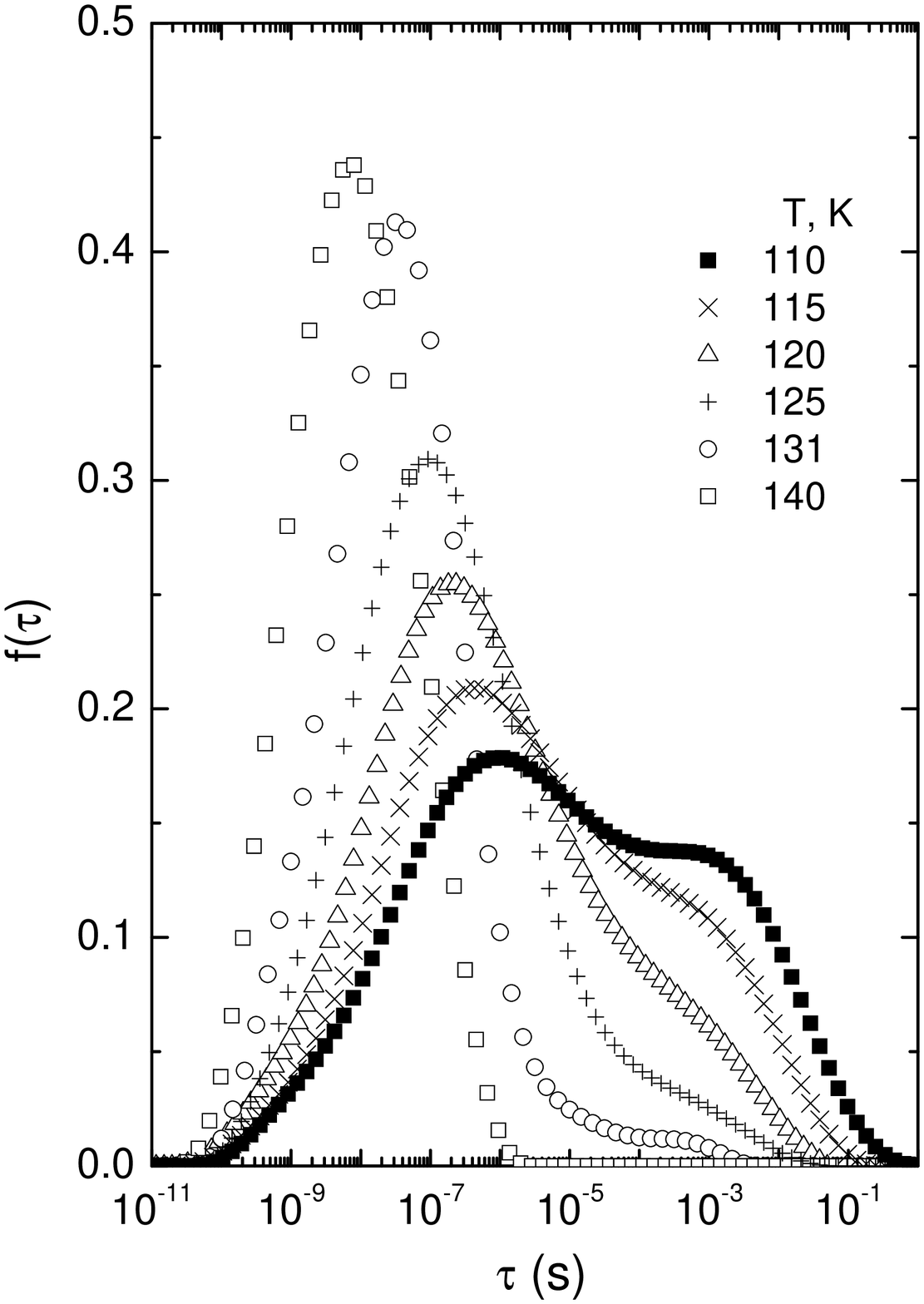}
 \end{center}
\caption{
 Relaxation time distribution for CuInP$_2$(S$_{0.25}$Se$_{0.75}$)$_6$
    crystals at various temperatures.
    } \label{Fig_6}
\end{figure}
This is clearly visible in Fig. 5, where the Cole-Cole fit is shown
as doted line. Not only Cole-Cole formula, however, other very well
known predefined dielectric dispersion formulas, such as
Havriliak-Negami, Cole-Davidson cannot adequate describe the
dielectric dispersion of the presented crystals. More general approach
must be used for determination of the broad continuous
distribution function of relaxation times $f(\tau)$ by solving a
Fredholm integral equations
\begin{subequations}\label{integr}
\begin{eqnarray}
\label{integ1}
    \varepsilon'(\omega)=\varepsilon_{\infty}+\Delta\varepsilon \int_{-\infty}^{\infty}\frac{f(\tau)d(ln\tau)}{1+\omega^2\tau^{2}},\\
\label{integ2}
    \varepsilon"(\omega)=\Delta\varepsilon \int_{-\infty}^{\infty}\frac{\omega\tau f(\tau)d(ln\tau)}{1+\omega^2\tau^{2}}.
\end{eqnarray}
\end{subequations}
with the normalization condition
\begin{equation}\label{normalization}
\int_{-\infty}^{\infty}f(\tau)d(ln\tau)=1.
\end{equation}
The most general method for the solution is the Tikhonov
regularization \cite{tikhonov, physrevst} method. The calculated
distribution of relaxation times of
CuInP$_2$(S$_{0.25}$Se$_{0.75}$)$_6$ crystals is presented in
Fig. 6. The symmetric and narrow distribution is observed only at
higher temperature \textit{T} ${>>}$ \textit{T$_m$} (at 1 kHz), on cooling the
distributions becomes broader and more asymmetric so that below
\textit{T$_m$} (at 1 kHz) second maximum appears. Such behaviour of
distribution of relaxation times have been already observed in a
very well known relaxors: Pb(Mg$_{1/3}$Nb$_{2/3}$)O$_3$ (PMN) \cite{grigalaitis1}, Pb(Mg$_{1/3}$Nb$_{2/3}$)O$_3$-Pb(Zn$_{1/3}$Nb$_{2/3}$)O$_3$-Pb(Sc$_{1/2}$Nb$_{1/2}$)O$_3$ (PMN-PZN-PSN)
\cite{macutkevic1},  Pb(Mg$_{1/3}$Ta$_{2/3}$)O$_3$ (PMT) \cite{kamba} and Sr$_{0.61}$Ba$_{0.39}$Nb$_2$O$_6$ (SBN) \cite{banys3}. From calculated
distributions of relaxation times the most probable relaxation
time ${\tau}$$_{mp}$, longest relaxation time ${\tau}$$_{max}$ and
${\tau}$$_{min}$ shortest relaxation time (the level 0.1 was
chosen as sufficient accurate) has been obtained (Fig. 7). The
shortest relaxation time ${\tau}$$_{min}$ is about 0.1 ns for
CuInP$_2$(S$_{0.25}$Se$_{0.75}$)$_6$ and about 0.01 ns for
CuInP$_2$(S$_{0.2}$Se$_{0.8}$)$_6$; it increases slowly with the
increase of temperature. The longest relaxation time
${\tau}$$_{max}$ diverges according to the Vogel-Fulcher law
\begin{equation}\label{Vogel}
\tau_{max}=\tau_{0max}\exp{\frac{E_{max}}{k(T-T_{0})}},
\end{equation}
where \textit{T$_{0}$} is the freezing temperature, \textit{E$_{max}$} is the activation energy of the longest relaxation times ${\tau}$$_{max}$ and $\tau$$_{0max}$ is the longest relaxation time at very high temperatures. 
The obtained parameters are presented in Table~\ref{table3},
however the most probable relaxation time ${\tau}$$_{mp}$ diverges with good
accuracy according to the Arrhenius law:
\begin{equation}\label{Arrhenius}
\tau_{mp}=\tau_{0mp}\exp{\frac{E_{mp}}{kT}},
\end{equation}
where \textit{E$_{mp}$} is the activation energy of the most probable relaxation times ${\tau}$$_{mp}$, and $\tau$$_{0mp}$ is the most probable relaxation time at very high temperatures. Obtained parameters are ${\tau}$$_{0mp}$ = 4.6$\times$10$^{-16}$ s and \textit{E$_{mp}$}/k
= 2365.3 K for CuInP$_2$(Se$_{0.75}$S$_{0.25}$)$_6$ and
${\tau}$$_{0mp}$ = 1.2$\times$10$^{-14}$ s and \textit{E$_{mp}$}/k = 1806.3 K for
CuInP$_2$(Se$_{0.8}$S$_{0.2}$)$_6$.
\begin{table}
\caption{\label{table3} Parameters of the Vogel-Fulcher fit of the
temperature dependencies of the longest relaxation times
$\tau_{max}$ in CuInP$_2$(S$_x$Se$_{1-x}$)$_6$ crystals with 0.2$\leq$x$\leq$0.25.}
\begin{ruledtabular}
\begin{tabular}{ccccc}
  compound & ${\tau}$$_{0max}$, s & \textit{T$_0$}, K & \textit{E$_{max}$}/k, K \\
\hline

\hline\\

  CuInP$_2$(Se$_{0.75}$S$_{0.25}$)$_6$  & 2.52$\times$10$^{-8}$  & 118.9  & 60.5 \\
  CuInP$_2$(Se$_{0.8}$S$_{0.2}$)$_6$  & 1.02$\times$10$^{-10}$  & 129.4 & 211.01 \\
           \end{tabular}
 \end{ruledtabular}
\end{table}
\begin{figure}
\begin{center}
    \includegraphics[width=85mm]{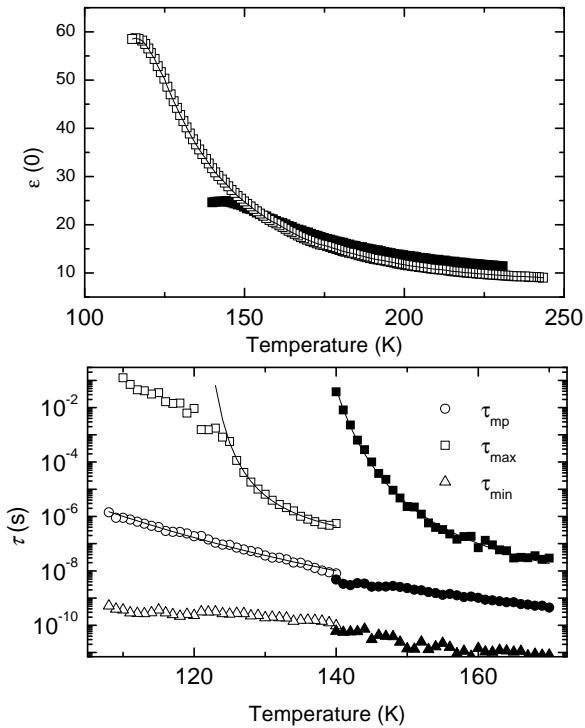}
  \end{center}
\caption{ Temperature dependence of the longest $\tau_{max}$,
most probable $\tau_{mp}$, shortest $\tau_{min}$ relaxation
times and static dielectric permittivity $\varepsilon$(0) in CuInP$_2$(S$_{x}$Se$_{1-x}$)$_6$
    crystals, with \textit{x}=0.2 (solid points) and \textit{x}=0.25 (open points). The
$\tau$(\textit{T}) lines were obtained from Vogel-Fulcher (for longest
relaxation times) and from Arrenius (for most probable relaxation
times) fits. The static dielectric permittivity
$\varepsilon$(0) lines were obtained from Eqs.
~\ref{Blinc} and ~\ref{Edwards}.}
\label{Fig_7}
\end{figure}
Such phenomenon can be caused by a distribution of Vogel-Fulcher
temperatures \textit{T$_0$}, where 0$\leq$\textit{T$_0$}$\leq$T$^{max}$$_0$
\cite{pirc}, \cite{pirc3}. In our case \textit{T$^{max}$} would correspond to a
Vogel-Fulcher temperature of $\tau_{max}$ and 0 is the freezing
temperature of the most probable relaxation time and all shorter
relaxation times. The temperature dependence of the reciprocal
static dielectric permittivity 1/$\varepsilon$(0) was fitted with
sperical random bond random field (SRBRF)
\begin{equation}\label{Blinc}
\varepsilon(0)=\frac{C_p(1-q_{EA})}{kT-J(1-q_{EA})},
\end{equation}
where \textit{J} is the mean coupling constant and \textit{q$_{EA}$} is
Edwards-Anderson order parameter, if \textit{q$_{EA}$}=0 then this equation
becomes the Curie-Weiss law.  The Edwards-Anderson order parameter
\textit{q$_{EA}$} for relaxor can be determined by equation \cite{blinc}:
\begin{equation}\label{Edwards}
q_{EA}=(\frac{\Delta J}{kT})^2(q_{EA}+\frac{\Delta f}{(\Delta
J)^2})(1-q_{EA})^2,
\end{equation}
where \textit{$\Delta$J} is the variance of the coupling and \textit{$\Delta$f} is
the variance of the random fields. Obtained parameters we will
discussed further together with random bonds random fields
parameters of other mixed crystals. We must admit that the
equations of the SRBRF model describe well static dielectric
properties of the presented crystals. At sulphur concentrations
between \textit{x}=0.25 and \textit{x}=0.2,
morphotropic phase boundary between the paraelectric
phases C2/c (characteristic for CuInP$_2$S$_6$) and P-31c
(characteristic for CuInP$_2$Se$_6$) or respectively ferrielectric
phases Cc and P31c were suggested \cite{vysochanskii4}. These results were later confirmed by X-ray and Raman investigations \cite{vysochanskii5}. Therefore, the disorder in these mixed crystals
is very high, and it can be reason of relaxor nature of the presented
crystals.

\subsection{\label{sec:levelC} Dipolar glass phase in mixed CuInP$_2$(S$_x$Se$_{1-x}$)$_6$ crystals }
For CuInP$_2$(S$_x$Se$_{1-x}$)$_6$ crystals with x=0.4-0.9 no
anomaly in static dielectric permittivity indicating the polar
phase transition can be detected down to the lowest temperatures.
The dielectric spectra of these crystals are very similar. As an
example, real and imaginary parts of the complex dielectric
permittivity of CuInP$_2$(S$_{0.8}$Se$_{0.2}$)$_6$ crystals are
shown in Fig. 8 as a function of temperature at several
frequencies.
\begin{figure}
  \begin{center}
   \includegraphics[width=83mm]{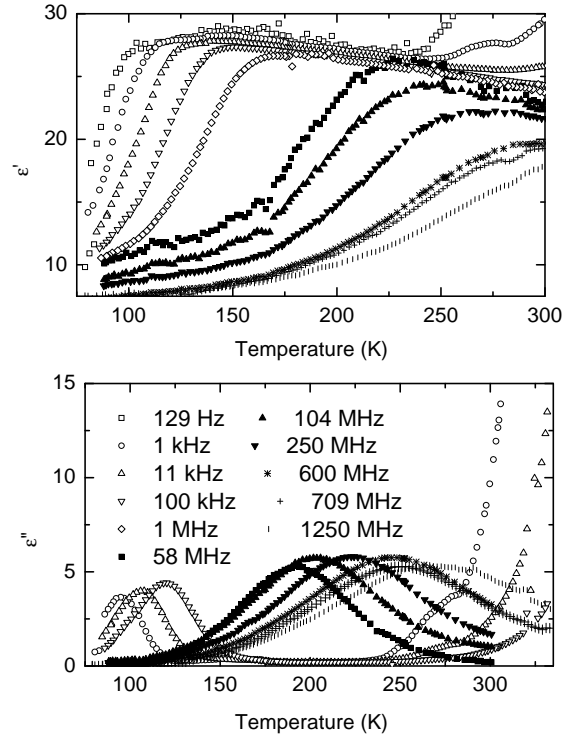}
  \end{center}
    \caption{
    Temperature dependence of the complex dielectric permittivity of CuInP$_2$(S$_{0.8}$Se$_{0.2}$)$_6$
    crystals measured at several
frequencies. }
    \label{Fig_8}
\end{figure}
It is easy to see a broad dispersion of the complex
dielectric permittivity starting from 260 K and extending to the
lowest temperatures. The maximum of the real part of dielectric
permittivity shifts to higher temperatures with increase of the
frequency together with the maximum of the imaginary part and
manifests typical behaviour of dipolar glasses. The dielectric
dispersion is symmetric of all crystals under study so that it can
easily be described by the Cole-Cole formula  (Fig. 9).
\begin{figure}
\begin{center}
    \includegraphics[width=83mm]{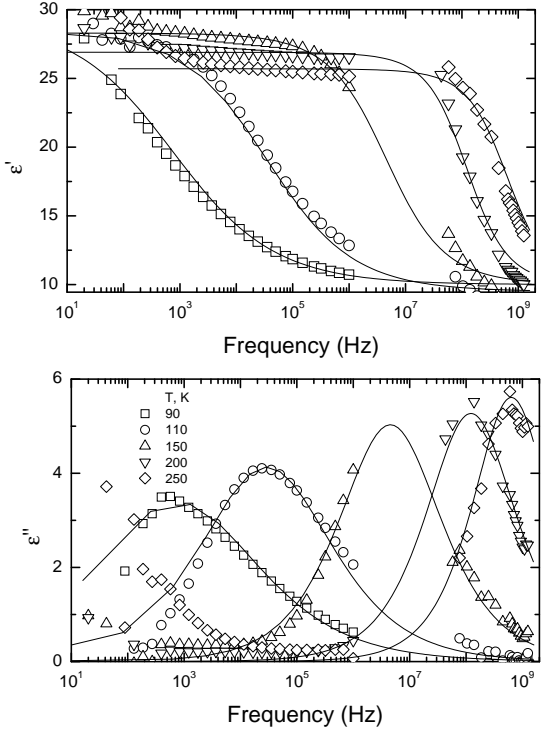}
  \end{center}
\caption{ Frequency dependence of the complex dielectric
permittivity of CuInP$_2$(S$_{0.8}$Se$_{0.2}$)$_6$
    crystals at several
temperatures. Lines are results of Cole-Cole fits. }
\label{Fig_9}
\end{figure}
The temperature dependence of the Cole-Cole parameters confirms
typical behaviour for dipolar glasses (Fig. 10): the mean Cole-Cole
relaxation time diverge according to the Vogel-Fulcher law (Eq.
~\ref{Vogel}), the Cole-Cole distribution parameter
$\alpha$$_{CC}$ strongly increases on cooling and reaches  value
0.5 below 100 K, the static dielectric permittivity temperature
dependence has no expressed maxima.
Usually such behaviour is analyzed in terms of the
three-dimensional random-bond random-field (3D RBRF) Ising model
of Pirc et al \cite{tadic}. In terms of this model, the temperature
dependence of static dielectric permittivity can be described with
the Eq. ~\ref{Blinc}. The order parameter is defined by the two
coupled self-consistent equations\cite{lopes}
\begin{equation}\label{Polarization}
P=\int_{-\infty}^{\infty}\frac{dz}{(2\pi)^{0.5}}tanh(\frac{\eta}{kT})exp{(-\frac{z^2}{2})},
\end{equation}

\begin{equation}\label{Anderson}
q_{EA}=\int_{-\infty}^{\infty}\frac{dz}{(2\pi)^{0.5}}tanh^2(\frac{
\eta}{kT})exp{(-\frac{z^2}{2})},
\end{equation}
where \textit{P} is the polarization and
\begin{equation}\label{eta}
\eta=(\Delta
J^2q_{EA}+\Delta f)^{0.5}z+JP.
\end{equation}
The Equation ~\ref{Blinc} describe good enough static dielectric
properties of presented dipolar glasses and obtained parameters
are in good agreement with parameters obtained from Vogel-Fulcher
fits, according to formula \cite{kind}
\begin{equation}\label{Freezing}
T_0=\Delta J/k_B.
\end{equation}
Obtained parameters we will discuss further below together with
random bonds random fields parameters of other mixed crystals.
\begin{figure} [h]
\begin{center}
    \includegraphics[width=85mm]{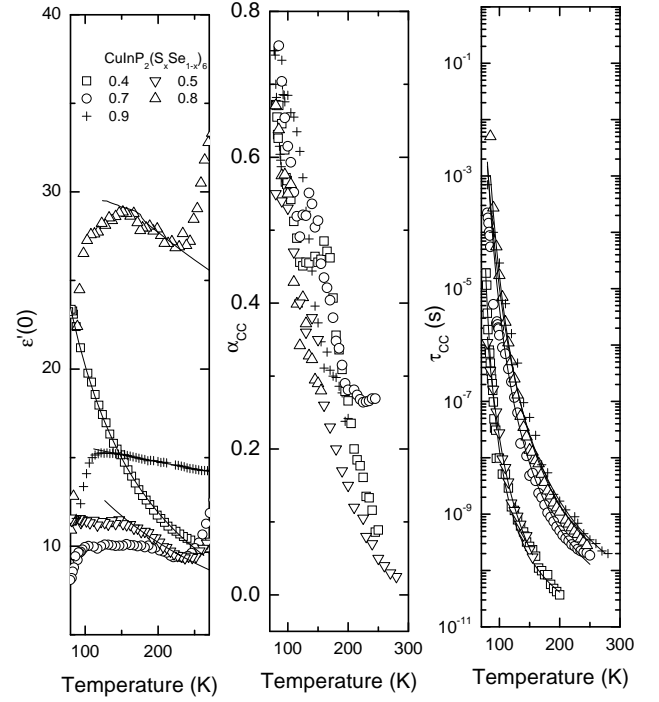}
  \end{center}
\caption{ Temperature dependence of the Cole-Cole parameters of
complex dielectric permittivity for the
CuInP$_2$(S$_{x}$Se$_{1-x}$)$_6$
    crystals with 0.4$\leq$x$\leq$0.9. The
 $\tau$ lines were obtained from Vogel-Fulcher fit and the $\varepsilon$(0) lines were obtained from 3D RBRF model
fit.}
\label{Fig_10}
\end{figure}
\subsection{\label{sec:levelD}  Influence of small amount of selenium to phase transition dynamics in
CuInP$_2$S$_6$ crystals} 
Temperature dependence of the dielectric
permittivity of CuInP$_2$S$_6$ crystals with a small amount of
selenium (x=0.98) is presented in Fig. 11. A small amount of
selenium changes dielectric properties of CuInP$_2$S$_6$ crystals
significantly: the temperature of the main dielectric anomaly
shift from about 315 to 289 K, the maximum value of the dielectric
permittivity $\varepsilon'$ significantly decreases from about 180
to 40 (at 1 MHz), at higher frequencies (from about 10 MHz) the
peak of dielectric permittivity becomes frequency- dependent in
CuInP$_2$(S$_{0.98}$Se$_{0.02}$)$_6$ crystals and a critical
slowing down disappears \cite{banys1}. 
An additional dielectric
dispersion appears at low frequencies and at low temperatures. The
CuInP$_2$(S$_{0.95}$Se$_{0.05}$)$_6$ crystals  exhibit
qualitatively similar dielectric anomaly with T$_c$ and
$\varepsilon'_{max}$ shifting to lower values.
\begin{figure}
  \begin{center}
   \includegraphics[width=83mm]{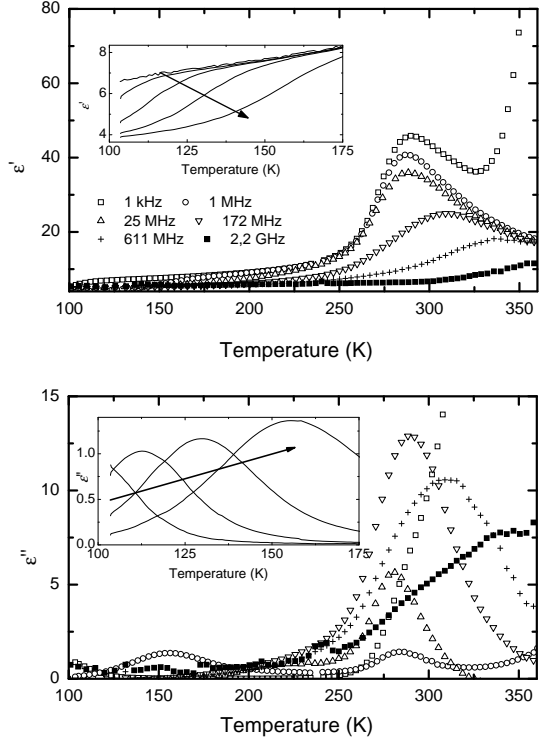}
  \end{center}
    \caption{
    Temperature dependence of the complex dielectric permittivity of CuInP$_2$(S$_{0.98}$Se$_{0.02}$)$_6$
    crystals measured at several
frequencies. }
    \label{Fig_11}
\end{figure}
\begin{figure}
  \begin{center}
   \includegraphics[width=83mm]{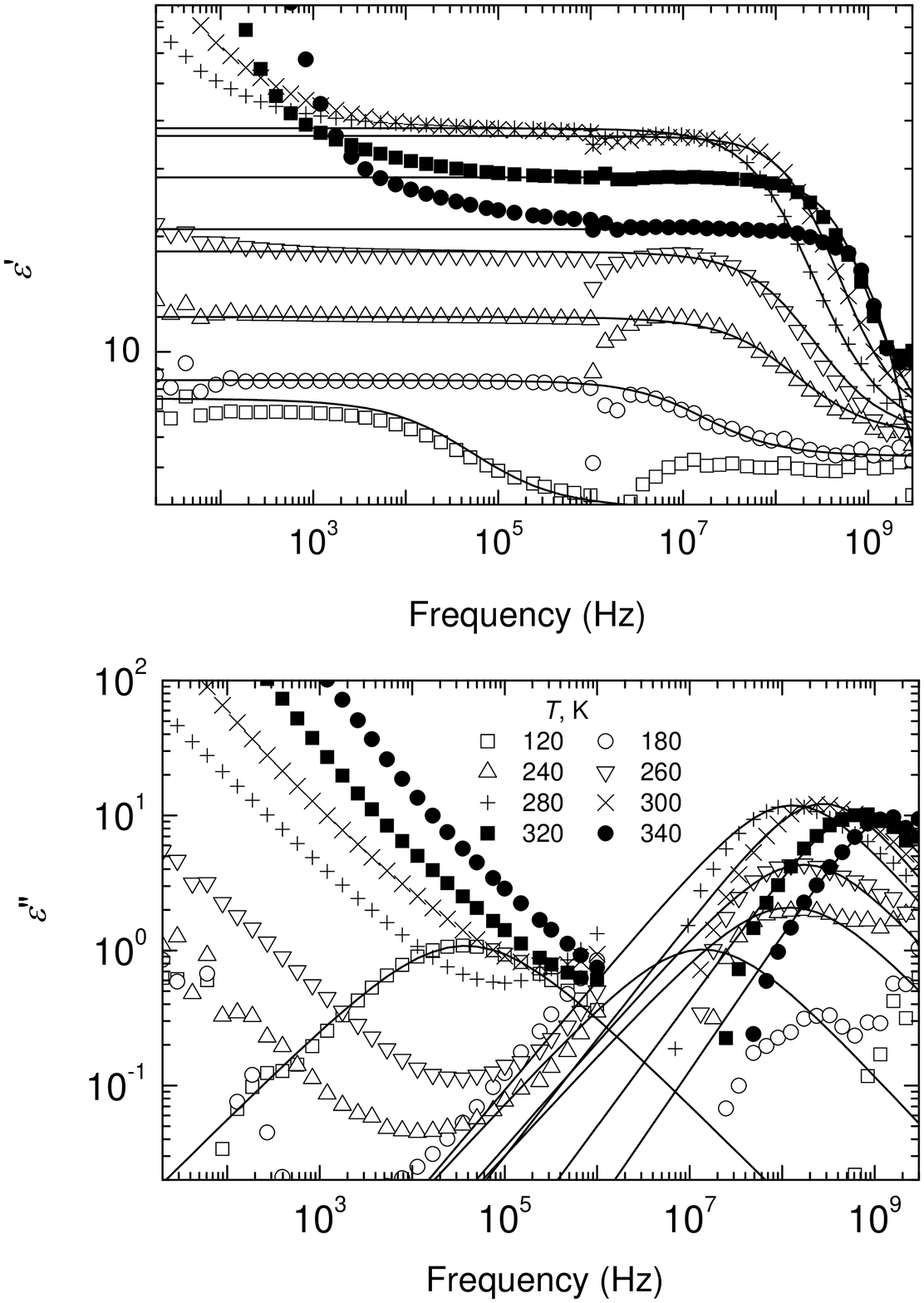}
  \end{center}
    \caption{
    Frequency dependence of the complex dielectric permittivity of CuInP$_2$(S$_{0.98}$Se$_{0.02}$)$_6$
    crystals measured at several
temperatures.Lines are results of Cole-Cole fits. }
    \label{Fig_12}
\end{figure}
The dielectric dispersion of presented crystals is symmetric (Fig. 12) so
that it can be correctly described by the Cole-Cole formula (Eq.
~\ref{Cole}). The Cole-Cole parameters are shown in Fig. 13. The
parameters of the Cole-Cole distribution  of relaxation
$\alpha$$_{CC}$ strongly increase on cooling and reach 0.43 at low
temperatures.
\begin{figure}
\begin{center}
    \includegraphics[width=85mm]{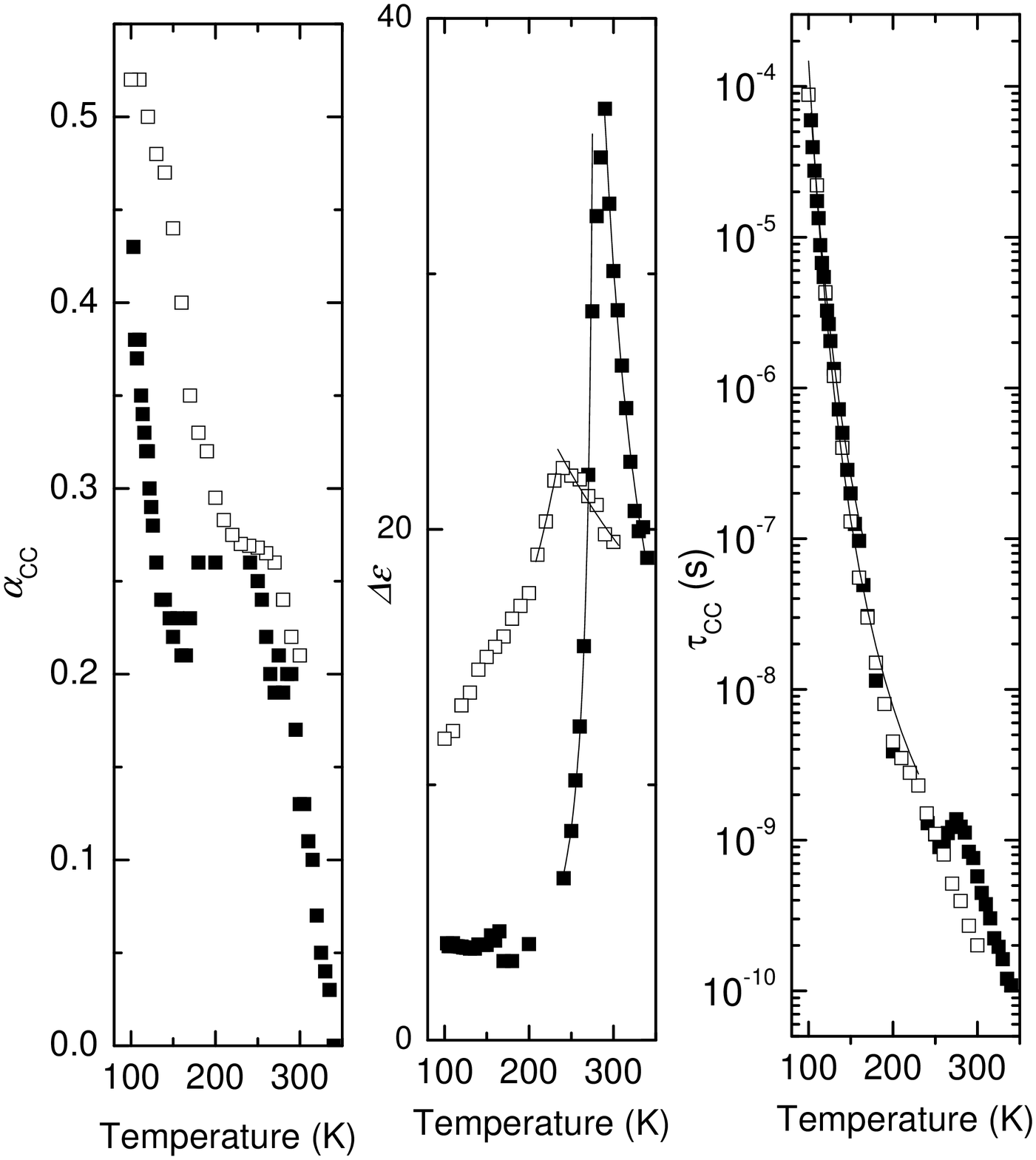}
 \end{center}
\caption{ Temperature dependence of the Cole-Cole parameters of
complex dielectric permittivity for the
CuInP$_2$(S$_{x}$Se$_{1-x}$)$_6$
    crystals with x=0.95 (open points) and x=0.98 (solid points). The
 $\tau$ lines were obtained from Vogel-Fulcher fit and the $\Delta$$\varepsilon$ lines were obtained from Curie-Weiss
fit. }
\label{Fig_13}
\end{figure}
The temperature dependence of the dielectric
strength $\Delta\varepsilon$  was fitted with the Curie-Weiss law
(Eq. ~\ref{Ciuri}). Obtained parameters are summarized in Table
~\ref{table4}.
\begin{table}
\caption{\label{table4} Parameters of phase transition dynamic of
CuInP$_2$S$_6$ crystals with small admixture of selenium. }
\begin{ruledtabular}
\begin{tabular}{ccccc}
  compound & C$_p$, K & C$_p$/C$_f$ & T$_{Cp}$, K  & T$_{Cf}$, K \\
\hline

\hline\\

CuInP$_2$(Se$_{0.05}$S$_{0.95}$)$_6$ & 8587.7 & 2.99 & 137.2 & 368.7 \\
CuInP$_2$(Se$_{0.02}$S$_{0.98}$)$_6$  & 1906.5 & 7.01 & 236.9 & 282.6 \\
          \end{tabular}
 \end{ruledtabular}
\end{table}
The difference T$_{Cp}$-T$_{Cf}$ and ratio C$_p$/C$_f$ in these
crystals indicate a first order, order-disorder phase transition.
In ferroelectric phase the mean relaxation time $\tau$$_{CC}$
decreases only in a narrow temperature region and only for
CuInP$_2$(S$_{0.98}$Se$_{0.02}$)$_6$, further on cooling a
significant increasing of times $\tau$$_{CC}$ is observed. This
increasing can be easily explained by the Fogel-Vulcher law (Eq.
~\ref{Vogel}). These parameters are summarized in Table
~\ref{table5}.
\begin{table}
\caption{\label{table5} Parameters of the Vogel-Fulcher fit of the
temperature dependencies of the mean relaxation times $\tau_{CC}$
in CuInP$_2$(S$_x$Se$_{1-x}$)$_6$ inhomogeneous ferroelectrics. }
\begin{ruledtabular}
\begin{tabular}{ccccc}
  compound & ${\tau_0}$, s & T$_0$, K & E/k, K \\
\hline

\hline\\

  CuInP$_2$(Se$_{0.95}$S$_{0.05}$)$_6$  & 8.5$\times$10$^{-12}$  & 1150  & 31 \\
  CuInP$_2$(Se$_{0.98}$S$_{0.02}$)$_6$  & 3.77$\times$10$^{-11}$  & 1215 & 28 \\
           \end{tabular}
 \end{ruledtabular}
\end{table}
Note that all parameters of different compounds in Table
~\ref{table5} are close to each other. Such a behaviour is very similar
to behaviour of betaine phosphite with a small amount of betaine
phosphate \cite{banys4} and in RADA \cite{trybula} crystals, where
a proposition that a coexistence of the
ferroelectric order and dipolar glass disorder appears at low temperatures was
proposed. Therefore we can conclude that mixed
CuInP$_2$(S$_x$Se$_{1-x}$)$_6$  crystals with x$\geq$0.95 also
exhibit at low temperatures a coexistence of ferroelectric and
dipolar glass disorder.
\subsection{\label{sec:levelE}  Phase diagram}
In this section we will discuss phase diagram in terms of random
bonds and random fields. For ferroelectrics we assume that mean
coupling constant \textit{J}/k is equal to \textit{T$_C$}, because Curie-Weiss fit
is accurate for these compounds and in this case Eq. ~\ref{Blinc} becomes Curie-Weiss law. Also for crystals
with with x$\leq$0.1, for the same reason we assume that $\Delta$J
and \textit{$\Delta$f} are 0. For ferroelectrics with \textit{x}$\geq$0.95 we
obtained \textit{$\Delta$J} from \textit{T$_0$} (Eq. ~\ref{Freezing}), we assumed
that \textit{$\Delta$f}=0.

\begin{figure}[t]
\begin{center}
    \includegraphics[width=85mm]{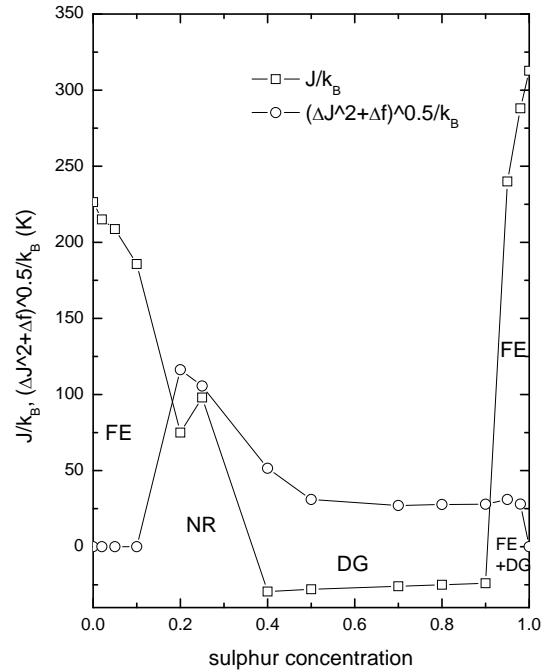}
  \end{center}
\caption{ Phase diagram of the mixed
CuInP$_2$(S$_{x}$Se$_{1-x}$)$_6$
    crystals (FE - ferroelectric phase, NR - nonergodic relaxor phase, DG - dipolar glass phase, FE+DG - ferroelectric and dipolar glass coexistence).
    } \label{Fig_14}
\end{figure}
In Fig. 14 we present the obtained phase diagram of mixed
crystals. In the mixed CuInP$_2$(S$_x$Se$_{1-x}$)$_6$ with
\textit{x}$\geq$0.95 and  \textit{x}$\leq$0.1 crystals the mean coupling constant
\textit{J}$>$(\textit{$\Delta$f}+\textit{$\Delta$J$^2$)$^{0.5}$}, therefore, they undergo
ferroelectric phase transition at \textit{J}/k. However is significant
difference between phase transition dynamics of mixed crystals
with \textit{x}$\geq$0.95 and \textit{x}$\leq$0.1. In mixed crystals with \textit{x}$\leq$0.1
no any coexistence of ferroelectric order and dipolar glass
disorder is observed down to the lowest temperature (80 K). At temperatures below 100 K the dielectric permittivity of these
compounds is very low (about 3), therefore, the phase coexistence
in these compounds is unlikely. In the ferroelectric phase these
crystals split into domains, it is evidenced in low
frequency dielectric dispersion spectra (Fig.1). However similar
ferroelectric domains already are observed in pure CuInP$_2$Se$_6$
crystals \cite{samulionis}. Really, influence of small amount of
sulphur to phase transition dynamics of mixed crystals appears
only by reduction \textit{T$_C$} (Table ~\ref{table1}). The influence of
small amount of selenium to phase transition dynamics is more
significant - already at x=0.95 the ferroelectric phase transition
in $\tau_{CC}$ is less expressed (Fig. 13). Such influence is
expressed also in other properties: rapid decreasing in \textit{T$_C$},
appearance of ferroelectric and dipolar glass phase coexistence at
\textit{x}=0.98 and onset of dipolar glass disorder with \textit{x} between 0.9 and
0.95.

For crystals with \textit{x}=0.2 and 0.25
\textit{J}$<$(\textit{$\Delta$f}+\textit{$\Delta$J}$^2$)$^{0.5}$ and  \textit{J}$\approx$
(\textit{$\Delta$f}+\textit{$\Delta$J}$^2$)$^{0.5}$ therefore the nonergodic relaxor
phase appears in these crystals at low temperatures. In the
presence of an external electric field \textit{E} meaning coupling constant
\textit{J} is expected to vary as
\begin{equation}\label{external}
J(E)=J(0)+\alpha E^2.
\end{equation}
For electrical field \textit{E} that \textit{J}(\textit{E}) $>$
(\textit{$\Delta$f}+\textit{$\Delta$J}$^2$)$^{0.5}$, in mixed crystals should be
observed relaxor to ferroelectric phase transition. The possible
existence of relaxor phase in mixed
ferroelectric-antiferroelectric crystals is stated in
\cite{matsuhita1,korner}. Really, no any evidence is indicated for polar nanoregions existence in mixed crystals. We try to fill this gap of information presenting  two mixed crystals,
where dielectric behaviour is very similar to very well known
relaxors PMN \cite{levstik} and SBN \cite{kleeman} (the
differencies are only in \textit{T$_m$} and $\varepsilon'_{max}$ values).
On the other hand, in phase diagram with less selenium
concentration no area with nonergodic relaxor phase (Fig. 14)
appears.  The main cause of such phase diagram is that disorder
((\textit{$\Delta$f}+\textit{$\Delta$J}$^2$)$^{0.5}$) is highest at \textit{x}=0.2, where
mean coupling constant is also high enough. Usually, for mixed
crystals is assumed that concentration dependence for \textit{$\Delta$f} is
such \cite{tadic}
\begin{equation}\label{mixed1}
\Delta f=4x(1-x)\Delta f_{max}.
\end{equation}
For \textit{$\Delta$J} similar behaviour also was assumed. In this case if
\textit{J} has minimum at x=0.5 the nonergodic relaxor phase can not be
observed. However any existing theories can not explain \textit{$\Delta$J}
and \textit{$\Delta$f} concentration dependence.

For compounds  0.9$\geq$x$\geq$0.4 the relation \textit{J}$\ll$
(\textit{$\Delta$f}+\textit{$\Delta$J}$^2$)$^{0.5}$ is valid, consequently in these compounds
a dipolar glass phase appears at low temperatures.
\section{\label{sec:level4}  Conclusions}
 The ferroelectric order in CuInP$_2$S$_6$  is reduced already for small (x=0.98) substitution of sulphur by selenium. By further increasing selenium concentration the dipolar glass phase appears. In contrast to in CuInP$_2$Se$_6$  even a high concentracion of admixture of sulphur (\textit{x}=0.1) has no any influence to the feroelectric order. The some degree of ferroelectric order exist even for \textit{x}=0.2 and \textit{x}=0.25, however, in these crystals the ferroelectricity is broken into polar nano regions. The random bonds and random fields model clearly describe the asymmetricity of phase diagram of mixed CuInP$_2$(S$_x$Se$_{1-x}$)$_6$, however this model can not identified origin of the effect. To summarize, the first experimental evidence for smearing nonergodic relaxor phase into dipolar glass phase by some doping is presented. For other relaxors the search of some admixture which transforms relaxor state into dipolar glass can also be performed.

\end{document}